\documentclass[galaxies,article,submit,pdftex,moreauthors]{mdpi}
\usepackage{natbib}
\newcommand{\ym}{\widetilde{y}}
\firstpage{1} 
\pubvolume{1}
\issuenum{1}
\articlenumber{0}
\pubyear{2022}
\copyrightyear{2022}
\datereceived{} 
\dateaccepted{} 
\datepublished{} 
\hreflink{https://doi.org/} 
\Title{Dark matter effects on the compact star properties}

\Author{H. C. Das $^{1,2}$\orcidA{}, Ankit Kumar $^{1,2}$\orcidB{}, Bharat Kumar $^{3}$\orcidC{}* and S. K. Patra $^{1,2}$\orcidD{}}

\AuthorNames{H. C. Das, A. Kumar, B. Kumar and S. K. Patra}

\AuthorCitation{Das, H. C.; Kumar, A.; Kumar, B.; Patra, S. K.}

\address{%
	$^{1}$ \quad Institute of Physics, Sachivalaya Marg, Bhubaneswar 751005, India; \\
	$^{2}$ \quad Homi Bhabha National Institute, Training School Complex, Anushakti Nagar, Mumbai 400094, India; \\
	$^{3}$ \quad Department of Physics \& Astronomy, National Institute of Technology, Rourkela 769008, India.}

\corres{Correspondence: kumarbh@nitrkl.ac.in}

\abstract{The neutron star properties are generally determined by the equation of state of $\beta$-equilibrated dense matter. In this work, we consider the interaction of fermionic dark matter (DM) particles with the nucleons via Higgs exchange and investigate its effect on the neutron star properties with the relativistic mean-field model equation of state coupled with DM. We deduce that DM significantly affects the neutron star properties, such as considerably reduce the maximum mass of the star, which depends on the percentage of the DM considered inside the neutron star. The tidal Love numbers both for electric and magnetic cases and surficial Love numbers are also studied for DM admixed NS. It is observed that the magnitude of tidal and surficial Love numbers increase with more DM percentage. Further, we point out that post-Newtonian tidal corrections to gravitational waves decreased by increasing DM percentage. Also, the DM effect on the GW signal is significant during the late inspiral and merger stages of binary evolution for GW frequencies >500 Hz.}

\keyword{Neutron Star; Dark Matter; Gravitational Waves.} 

\begin{document}
\section{Introduction}
	Gravitational waves (GWs) are emitted from binary neutron star (BNS) mergers, such as GW170817~\cite{Abbott_2017} and GW190425~\cite{Abbott_2020}. These are exceptional discoveries by advanced LIGO~\cite{LIGO} and Virgo \cite{Virgo} detectors. {The detected GWs, including the electromagnetic counterparts~\cite{Drout2017,Cowperthwaite_2017,Chornock_2017,Nicholl2017}, opened up a new era of multi-messenger astronomy---which helps us to study the properties of compact objects. By combining the total binary mass of GW170817 inferred from GW signal with electromagnetic observations, an upper limit of the neutron star (NS) mass $M_{{\rm max}}\leq2.17 \ M_\odot$ was predicted by Margalit et al. \cite{Margalit_2017}. 
		Rezzolla et al. presented the upper bound by combining the GW observations and quasiuniversal relations as $M_{{\rm max}}\leq 2.16_{-0.15}^{+0.17}\ M_\odot$ \cite{Rezzolla_2018}. Further analysis by employing both energy and momentum conservation laws and the numerical-relativity simulations showed that the maximum mass of a cold NS is bound to be less than $2.3\ M_\odot$ \cite{Shibata_2019}. 
		Various massive pulsar discoveries constrained the equation of states (EOSs) of the supra-nuclear matter inside the core of the NS \cite{Demorest_2010, Antoniadis_2013, Cromartie_2020}. These observational data also suggest strong constraints on the maximum mass of the slowly rotating NS with a lower bound of $\sim$$2\ M_\odot$, which discards many equation of states. Recently, the Neutron Star Interior Composition Explorer (NICER) data also imposed stringent constraints on the mass and radius of a canonical star from the analysis of PSR J0030+0451 data \cite{Miller_2019, Riley_2019, Raaijmakers_2019, Miller_2021}.} 
	
	GWs discoveries of BNS mergers can also be a novel probe for nuclear physics as they encode information about the EOS in neutron-rich dense matter through finite-size effects and also provide information on NS properties, such as the mass and tidal deformability \cite{Hinderer_2008,Hinderer_2010,Agathos2015,Fattoyev_2018,Tuhin_2018,Most_2018,Raithel_2018,Erik2019,Han2019,Fattoyev_2020,Niels2020,Andrea2020,Chamel_2020,Chamel_2021}. In the NS-NS binary, the tidal bulges of the stars affect each other, which draws energy out of the orbit and slightly accelerates the merger---indicating that the GW signal is phase-shifted relative to a merger of tideless point-masses. 
	The Love numbers that are indirectly influenced by the EOS can also be enumerated using the phase shift observation by {advanced LIGO \cite{LIGO}, Virgo \cite{Virgo} and KAGRA \cite{Kagra2020}} detectors.
		 As the NSs are not point masses, the tides are not purely quadrupolar, and we need octupoles, hexadecupoles, etc. Love numbers $(k_3, k_4,  $\ldots$)$  fully describe the deformation. In any case, the quadrupole gravitational Love numbers $k_2^{el}$ (electric-type) and $j_2^{mag}$ (magnetic-type) produce the dominant effect and are of the greatest interest to GW astrophysicists \cite{KumarTide_2017}. 
		 Tidal distortion enters the picture when the BNS are in close orbit with each other. Initial estimations suggest that the tidal effects give dominant contributions mainly in the late inspiral phase with the initial frequency of more than 300 Hz. To calculate the inspiral properties, we use post-Newtonian (PN) formalism and use the tidal corrections up to $(2l+1)$ PN order as given in Ref. \cite{Yagi_2014}. 
		 Although the PN method breaks down in the higher frequency region, it continues to work up to some extent. For that, one has to take the highest order PN order both for point-particle and tidal contributions available to date as given in Refs. \cite{Vines_2011}.  Further, gravitational waves from  BNS mergers also opened a new possibility for models of dark matter. Due to dark matter, the gravitational wave signal emitted from the NS merger could be modified with post-merger peaks \cite{Ellis_2018}. These GW peaks could be detected in future GW signals from BNS mergers. In Refs. \cite{Das_2020, DasMNRAS_2021, DasPRD_2021}, we noticed that dark matter could affect the NS mass--radius relation and tidal deformability.
    
    We know that dark matter (DM) exists in the Universe as a whole to individual galaxies, including our Milky Way and, to some extent, to smaller  scale structures, but this is hard to detect and measure. The astrophysical observations are in the favour of the existence of DM. Thus, we know DM exists but do not know what DM is made of? To know its exact nature, a large number of  experiments along with various theoretical modelling are carried on; a concrete picture on this DM is not yet drawn. The popular DM candidates are the weakly interacting massive particles (WIMPs) \cite{Kouvaris_2011}, \mbox{Neutralino~\cite{Hooper_2004,Han_2014,Das_2019}}, feebly interacting massive particles (FIMPs) \cite{Bernal_2017,Hall_2010}, Axions \cite{Duffy_2009} etc. 
Recently, the~Cryogenic Dark Matter Search (CDMS) experiment at Fermi lab in Illinois \cite{CDMS_2013} and the Large Underground Xenon (LUX) experiment in South Dakota \cite{LUX_2014}  reported the results from their searches for DM candidate particles known as WIMPs. Both these experiments attempted to directly detect WIMPs because they scatter off the nuclei of a detector. The CDMS locates a signal that may be assigned to WIMPs. However, LUX noticed no confirmation of such DM particles with nominally better sensitivity. The Fermi large area telescopes and imaging air Cherenkov telescopes were also established to indirectly measure the presence of DM \cite{Conrad_2014}.
	
    Neutron stars are ``cosmic laboratories'', allowing us to study how DM behaves under extreme conditions that cannot be reproduced on Earth. As NSs are extremely dense, they may trap a certain amount of DM particles that pass through them. Hence, there is a certain probability of the accretion of DM due to its huge gravitational potential and immense baryonic density. After accretion of the DM, this affects  properties of the NS. If~the DM is self-annihilating in nature, it heats the NS and fastens the cooling rate \cite{Kouvaris_2008,Bhat_2019}, while the non-self-annihilating DM is deposited in the star core and influences the structural properties \cite{DasPRD_2021,De_Lavallaz_2010, Ciarcelluti_2011}. 
    The effects of bosonic DM on the gravitational wave signals were studied in Ref. \cite{Rafiel_2022}. In our recent works \cite{Das_2020,Das_2021}, we considered the fermionic DM, which interacts with nucleons via the Higgs mechanism. The effects of DM on the structure of NS were analysed. Taking into account the DM admixture in the nuclear EOS, we can suggest the secondary component of the GW190814 data as a massive NS with an admixture of DM \cite{DasPRD_2021}. We also found a delay in BNS merger in the presence of the DM admixture \cite{DasPRD_2021}. 
    In the present calculations, we use the DM admixed EOS as in Ref. \cite{Das_2020} and analyse the results of the higher-order tidal Love numbers (electric-type and magnetic-type) and surficial love numbers of the NSs. With the increased sensitivity of upcoming third-generation GW detectors and gravitational-wave detection of BNS mergers, that analysis is potentially important \cite{Rana14,Etelescope}. 
    
    We organized this paper as follows: In Section \ref{formalism}, we present the formalism related to the  NS structure with DM accreted inside it. In Section \ref{R&D}, we discuss the results. Finally, the work is summarised in Section \ref{conclusion}.
	\section{Formalism}
	\label{formalism}
	\subsection{Dark Matter Admixed Neutron Star EOS}
	\label{form:EOS_mr}
	{The EOS of the NS is calculated with the effective field theory motivated relativistic mean-field (E-RMF) model, where the nucleons interact with each other by exchanging different mesons. The model Lagrangian was constructed by including the self and cross-couplings between the important mesons terms up to the fourth-order \cite{Kumar_2017, Kumar_2018, Kumar_2020, Parmar_2021}. The~leptons (electrons and muons) contribution are also taken into account to maintain both $\beta$-equilibrium and charge neutrality condition \cite{NKGb_1997}. The energy density (${\cal E}_{NS}$) and pressure ($P_{NS}$) of the NS is given by \cite{Kumar_2018, Dasfmode_2021}
    \begin{eqnarray}
    {\cal E}_{NS}& = & \sum_{i=n,p} \frac{\gamma_{n}}{2\pi^2}\int_{0}^{k_{F}} k^2\ dk \sqrt{k^2+m_n^{*2}}+n g_{\omega}\omega_0+\frac{n_{3}}{2}g_{\rho}\rho_0-\frac{1}{3!}\zeta_{0}{g_{\omega}^2}\omega_0^4
    \nonumber\\
    &+&m_{\sigma}^2{\sigma_0}^2\Bigg(\frac{1}{2}+\frac{\kappa_{3}}{3!}\frac{g_{\sigma}\sigma_0}{m_n}+\frac{\kappa_4}{4!}\frac{g_{\sigma}^2\sigma_0^2}{m_n^2}\Bigg)
    -\frac{1}{2}m_{\omega}^2\omega_0^2\Bigg(1+\eta_{1}\frac{g_{\sigma}\sigma_0}{m_n}+\frac{\eta_{2}}{2}\frac{g_{\sigma}^2\sigma_0^2}{m_n^2}\Bigg)
    \nonumber\\
    &-&\frac{1}{2}\Bigg(1+\frac{\eta_{\rho}g_{\sigma}\sigma_0}{m_n}\Bigg)m_{\rho}^2\rho_{03}^{2}-\Lambda_{\omega}g_{\rho}^2g_{\omega}^2\rho_{03}^2\omega_0^2
    \nonumber\\
    &+&\sum_l\frac{\gamma_{_l}}{2\pi^2}\int_0^{k_{F{_l}}} dk \ \sqrt{k^2+m_l^2},
    \label{eq:ENS}
    \end{eqnarray}
    and
    \begin{eqnarray}
    P_{NS}& = & \sum_{i=n,p} \frac{\gamma_n}{6\pi^2}\int_{0}^{k_F} \frac{k^4\ dk}{\sqrt{k^2+m_n^{*2}}} +\frac{1}{3!}\zeta_{0}{g_{\omega}^2}\omega_0^4+\Lambda_{\omega}g_{\rho}^2g_{\omega}^2\rho_{03}^2\omega_0^2
    \nonumber\\
    &-&m_{\sigma}^2{\sigma_0}^2\Bigg(\frac{1}{2}+\frac{\kappa_{3}}{3!}\frac{g_{\sigma}\sigma_0}{m_n}+\frac{\kappa_4}{4!}\frac{g_{\sigma}^2\sigma_0^2}{m_n^2}\Bigg)
    +\frac{1}{2}m_{\omega}^2\omega_0^2\Bigg(1+\eta_{1}\frac{g_{\sigma}\sigma_0}{m_n}+\frac{\eta_{2}}{2}\frac{g_{\sigma}^2\sigma_0^2}{m_n^2}\Bigg)
    \nonumber\\
    &+&\frac{1}{2}\Bigg(1+\frac{\eta_{\rho}g_{\sigma}\sigma_0}{m_n}\Bigg) m_{\rho}^2\rho_{03}^{2}
    +\sum_l\frac{\gamma_{_l}}{6\pi^2}\int_0^{k_{F{_l}}} \frac{k^2 dk}{\sqrt{k^2+m_l^2}},
    \label{eq:PNS}
    \end{eqnarray}
    where $\gamma_n$ and $\gamma_{l}$ are the spin degeneracy factor for nucleons and leptons, respectively. The~$k_F$ and $k_{F_l}$ are the nucleons and leptons Fermi momenta, respectively. All the coupling constants, masses and fields of different mesons are already given in Refs. \cite{Kumar_2018, Kumar_2020, Dasfmode_2021}. The~$m^*_n$ is the effective masses of the nucleons, which is written as
    \begin{eqnarray}
    m^*_n&=& m_n-g_{\sigma}\sigma_0-g_{\delta}\tau\delta_0.
    \label{eq:effm}
    \end{eqnarray} }
\vspace{-3pt}
	{Dark matter particles are believed to be accreted inside the NS, due to the huge gravitational potential and immense baryonic density. The accreted DM particles interact with baryons by exchanging standard model Higgs. The interacting Lagrangian is given in the form ~\cite{Panotopoulos_2017, Das_2019, Quddus_2020, Das_2020, Das_2021, DasMNRAS_2021, DasPRD_2021}:
    \begin{eqnarray}
    {\cal{L}}_{DM}&=&  \bar \chi \Big[ i \gamma^\mu \partial_\mu - m_\chi + y h \Big] \chi +  \frac{1}{2}\partial_\mu h \partial^\mu h 
     - \frac{1}{2} m_h^2 h^2 +\sum \frac{f m_n}{v} \bar{\psi} h \psi , 
    \label{eq:ldm}
    \end{eqnarray}
    where $\psi$ and $\chi$ are the baryons and DM wave functions respectively. The parameters $y$ is DM-Higgs coupling, $f$ is the nucleon-Higgs form factors  equal to 0.35, and $v$ is the vacuum expectation value of Higgs field. The values of $y$ and $v$ are 0.07 and 246 GeV, respectively taken from the Refs. ~\cite{Das_2020, Das_2021}. The  $m_n$, $m_\chi$ and $m_h$ are the masses of the nucleon, DM, and Higgs equal to 939 MeV, 200 GeV and 125 GeV, respectively.}
    {We calculate the spin-independent scattering cross-section of the nucleons with DM using the relation ~\cite{Cline_2013}
    \begin{eqnarray}
    \sigma_{SI}=\frac{y^2f^2m_n^2}{4\pi}\frac{\mu_r}{v^2m_h^2},
    \end{eqnarray}
    where $\mu_r$ is the reduced mass. The calculated cross-sections for $m_\chi=200$ GeV is found to be $9.70\times10^{-46}$ cm$^{2}$, which is  consistent with XENON-1T ~\cite{Xenon1T_2016}, PandaX-II ~\cite{PandaX_2016}, PandaX-4T \cite{Meng_2021} and LUX ~\cite{LUX_2017} within a 90\% confidence level. The LHC  also provided a limit on the WIMP-nucleon scattering cross-section in the range from $10^{-40}$ to $10^{-50}$ cm$^{2}$~~\cite{Djouadi_2012}. Thus, our model also satisfies the LHC limit. Therefore, in the present calculations, we constrained the value of $y$ from both the direct detection experiments and the LHC results.}

    {The nucleon-Higgs form factor $f$ was calculated in Ref. ~\cite{Djouadi_2012} using the implication of both lattice QCD ~\cite{Czarnecki_2010} and MILC results ~\cite{MILC_2009} whose value is $0.33_{-0.07}^{+0.30}$ ~\cite{Aad_2015}. The taken value of $f$ (=0.35) in this calculation lies in the region. Thus, we also constrain the values of $f$ with available data.
    The energy density (${\cal{E}}_{DM}$) and pressure ($P_{DM}$) for NS with DM can be obtained by solving  Equation (\ref{eq:ldm})
    \begin{eqnarray}
    {\cal{E}}_{DM}& = & \frac{1}{\pi^2}\int_0^{k_f^{DM}} k^2 \ dk \sqrt{k^2 + (m_\chi^\star)^2 } +\frac{1}{2}m_h^2 h_0^2 ,
    \label{eq:edm}
    \end{eqnarray}
    and
    \begin{eqnarray}
    P_{DM}& = & \frac{1}{3\pi^2}\int_0^{k_f^{DM}} \frac{ k^4 \ dk} {\sqrt{k^2 + (m_\chi^\star)^2}} - \frac{1}{2}m_h^2 h_0^2 ,
    \label{eq:pdm}
    \end{eqnarray} 
    $h_0$ is the Higgs field calculated by applying the mean-field approximation ~\cite{Das_2019}. $k_f^{DM}$ is the Fermi momentum for DM. The contribution of the Higgs field in both energy density and pressure is minimal. The $m_\chi^{\star}$ is the DM effective mass, which can be given as
    \begin{equation}
    m_{\chi}^{\star} {}=m_{\chi}-yh_0. 
    \label{eq:m_chi_star}
    \end{equation}}
%
    {Assuming the average number density of nucleons ($n_b$) is $10^3$-times larger than the average dark matter density ($n_{DM}$), this implies the ratio  of the dark matter and the NS mass to be $\sim$$\frac{1}{6}$ \cite{Panotopoulos_2017}. Hence, the value of $k_{f}^{DM}$ is $\sim$$0.033$ GeV. Therefore, we can vary DM Fermi momenta from 0 to 0.05 GeV to study the NS properties. The total EOS of the NS is the addition of the NS part, and DM is written as \cite{DasMNRAS_2021}
	\begin{eqnarray}
	{\cal E}  = {\cal E}_{NS} + {\cal E}_{DM}, \ {\rm and} \  P = P_{NS} + P_{DM}.
	\label{eq:total_EOS}
	\end{eqnarray}}
%
It was reported in our earlier works  \cite{Das_2020, Das_2021, DasPRD_2021, Dasfmode_2021,DasMNRAS_2021} that the  addition of DM in the NS changes  its structural properties. Here, we study the effects of DM on the gravitoelectric and gravitomagnetic tidal properties. In addition, we are also analyse the tidal correction on the GWs emitted by BNS in the in-spiral phase.
	
	\subsection{Mass and Radius of the NS}
	\label{form:mr}
	The mass and radius of the NS are calculated by solving the Tolman--Oppenheimer--Volkoff (TOV) equations, which are written as \cite{TOV1,TOV2}:
	\begin{eqnarray}
	\frac{dP(r)}{dr}&=&-\frac{[{\cal E}(r)+P(r)] [m(r)+{4\pi r^3P(r)}]}{r^2\Big(1-\frac{2m(r)}{r}\Big)},
	\nonumber \\
	\frac{dm(r)}{dr} &=& 4\pi r^2 {\cal E}(r).
	\label{eq:TOV}
	\end{eqnarray}
	Here, ${\cal E}(r)$ and $P(r)$ are the energy density and pressure, respectively. The enclosing mass $m(r)$ at a distance $r$ from the centre of the star is obtained by solving the TOV equations with boundary conditions $r=0$ and $P=P_c$ at a fixed central density. The maximum mass ($M$) of the NS and the corresponding radius ($R$) are obtained from the coupled differential equations assuming the pressure vanishes at the surface of the star, i.e., $P(r)=0$ at $r=R$.
	\subsection{Tidal Love Numbers and Tidal Deformability of the NS}
	\label{form:tidal}
	In a BNS system, when the two stars are in close orbit with each other, the tidal interactions come into the picture. The mutual gravitational interaction increases with time because the inspiralling stars emit gravitational radiation due to their acceleration in the gravitational field. The shape of the star becomes increasingly tidally deformed. Analogous to the electromagnetic radiation, here also, the gravitational field is generated from the two types of tidal fields, i.e., (i) electric (even parity or polar) ${\cal E}_L$ and (ii) magnetic (odd parity or axial) ${\cal M}_L$ fields,  where $L$ represents the space indices. These two fields induce a mass multipole moment ($Q_{L}$) and a current multipole moment ($S_{L}$), which are defined \mbox{as \cite{Damour_2009, Chamel_2021}}:
	\begin{equation}
	Q_L=\lambda_l {\cal E}_L,
	\label{eq:Q_L}
	\end{equation}
	\begin{equation}
	S_L=\sigma_l {\cal M}_L,
	\label{eq:S_L}
	\end{equation}
	where $\lambda_l$ and $\sigma_l$ are the gravitoelectric and gravitomagnetic tidal deformability of order $l$. Both the mass and current multipole moments come into the picture in general relativity as compared to the Newtonian dynamics, where only the mass multipole moment plays a role. The gravitoelectric ($k_l$) and gravitomagnetic Love numbers ($j_l$) are related with the deformability parameters $\lambda_l$ and $\sigma_l$, respectively, as \cite{Chamel_2021}:
	\begin{equation}
	k_l = \frac{(2l-1)!!}{2}\frac{\lambda_l}{R^{2l+1}},
	\label{eq:k_l}
	\end{equation}
	\begin{equation}
	j_l = 4(2l-1)!!\frac{\sigma_l}{R^{2l+1}}.
	\label{eq:j_l}
	\end{equation}
	The dimensionless tidal deformability parameters correspond to both electric $\Lambda_l$ and magnetic $\Sigma_l$ multipole moments, which are quite useful to express the tidal deformability of the BNSs. These quantities are often used conveniently in various measurements and related with the Love numbers and compactness of the star $C$ with the relation \cite{Chamel_2021}:
	\begin{equation}
	\Lambda_l=\frac{\lambda_l}{M^{2l+1}}=\frac{2}{(2l-1)!!}\frac{k_l}{C^{2l+1}}
	\label{eq:Lambda_l}
	\end{equation}
	\begin{equation}
	\Sigma_l= \frac{\sigma_l}{M^{2l+1}}=\frac{1}{4(2l-1)!!}\frac{j_l}{C^{2l+1}}.
	\label{eq:Sigma_l}
	\end{equation}
	
	As we  discussed earlier in this paper, the leading order perturbation of the tidal deformations is the gravitoelectric and gravitomagnetic quadrupole deformations $\lambda_2$ and $\sigma_2$. However, due to the close orbital separation of the binary system and also at the merging condition, the higher-order deformations are also quite substantial. All these tidal deformations can be expressed with the various Love numbers $k_l$ and $j_l$ as defined in Equations (\ref{eq:k_l}) and (\ref{eq:j_l}).  To calculate these tidal Love numbers along with the dimensionless tidal deformability $\Lambda_l$ and $\Sigma_l$, we adopt the formalism developed in \mbox{Ref. \cite{Chamel_2021}}. The gravitoelectric perturbations are evaluated from the differential equation in terms of metric functions ($H$), which are written as:
	\begin{eqnarray}
	H_\ell''(r)&+& H_\ell'(r) \biggl[1-\frac{2m(r)}{r}\biggr]^{-1} \Bigg\{\frac{2}{r} - \frac{2 m(r)}{r^2} - 4\pi r \left[{\cal E}(r) -  P(r)\right] \Bigg\} 
	\nonumber  \\
	&+& H_\ell(r) \biggl[1-\frac{2m(r)}{r}\biggr]^{-1} \times \Bigg\{4\pi \bigg[5{\cal E}(r)+ 9P(r) +\dfrac{d{\cal E}}{dP} \left[{\cal E}(r)  + P(r)\right] \bigg] 
	\nonumber  \\
	&-& \frac{\ell(\ell+1)}{r^2} - 4\biggl[1-\frac{2m(r)}{r}\biggr]^{-1} \biggl[\frac{m(r)}{r^2} + 4\pi\, r\,  P(r)\biggr]^2 \Bigg\} = 0, \;
	\label{eq:H_eq}
	\end{eqnarray}
	and the gravitomagnetic perturbations are obtained by solving the differential equation, which is given as:
	\begin{eqnarray}
	\widetilde{H}_{\ell}''(r)&-&\widetilde{H}_{\ell}'(r) \biggl[1-\frac{2 m(r)}{r}\biggr]^{-1} \times 4\pi r  \left[ P(r) + {\cal E} (r) \right]
	\nonumber  \\
	&-&\widetilde{H}_{\ell}(r) \biggl[1-\frac{2 m(r)}{r}\biggr]^{-1} \times \left\{ \frac{\ell(\ell+1)}{r^2} - \frac{4m(r)}{r^3} +8 \pi \theta  \left[ P(r) + {\cal E} (r) \right] \right\} = 0, 
	\label{eq:Ht_eq}
	\end{eqnarray} 
	with $\theta$ is $+1$ for static fluid and $\theta=-1$ is for irrotational fluid. These two second order differential equations (Equations (\ref{eq:H_eq}) and (\ref{eq:Ht_eq})) are solved by converting them to two first order coupled differential equations in terms of $\beta(r)=H_\ell'(r)$ and $\widetilde{\beta}(r)=\widetilde{H}_\ell'(r)$ respectively along  with the TOV equations [Equation (\ref{eq:TOV})]. The solutions of these differential equations at the surface of the star in terms of $y(R)$ and $\widetilde{y}(R)$ are \cite{Damour_2009, Chamel_2021}
	\begin{eqnarray}
	y(R)=\frac{R\beta(R)}{H_{\ell}(R)}, \ {\rm and} \ \widetilde{y}(R)=\frac{R\widetilde{\beta}(R)}{\widetilde{H}_{\ell}(R)}.
	\end{eqnarray}
	The gravitoelectric Love numbers $k_2$, $k_3$ and $k_4$ are given as \cite{Damour_2009, Hinderer_2008, Hinderer_2010, KumarTide_2017, Chamel_2021}:
	\begin{align}
	k_2 =& \, \frac{8}{5} C^5 (1-2C)^2 \big[ 2(y_2-1)C - y_2 + 2 \big]
	\nonumber \\ &
	\times \Big\{ 2C \big[ 4(y_2+1)C^4 + 2(3y_2-2)C^3 - 2(11y_2-13)C^2 + 3(5y_2-8)C - 3(y_2-2) \big]
	\nonumber \\ &
	+ 3(1-2C)^2 \big[ 2(y_2-1)C-y_2+2 \big] \log(1-2C) \Big\}^{-1} \, , 
	\label{eq:k2}
	\end{align}
	\begin{align} 
	&k_3 = \, \frac{8}{7} C^7 (1-2C)^2 \big[ 2(y_3-1)C^2 - 3(y_3-2)C + y_3 - 3 \big]
	\nonumber \\ &
	\times \Big\{ 2C \big[ 4(y_3+1)C^5 + 2(9y_3-2)C^4 - 20(7y_3-9)C^3 + 5(37y_3-72)C^2 - 45(2y_3-5)C 
	\nonumber \\ &
	+ 15(y_3-3) \big] + 15(1-2C)^2 \big[ 2(y_3-1)C^2 - 3(y_3-2)C + y_3 - 3 \big] \log(1-2C) \Big\}^{-1}  \, , 
	\label{eq:k3}
	\end{align}
	and
	\begin{align}
	&k_4 = \frac{32}{147} C^9 (1-2C)^2 \big[ 12(y_4-1)C^3 - 34(y_4-2)C^2 + 28(y_4-3)C - 7(y_4-4) \big]
	\nonumber \\
	&\times \Big\{ 2C \big[ 8(y_4+1)C^6 + 4(17y_4-2)C^5 - 12(83y_4-107)C^4 + 40(55y_4-116)C^3 \nonumber \\ &- 10(191y_4-536)C^2 + 105(7y_4-24)C - 105(y_4-4) \big] + 15(1-2C)^2 \big[ 12(y_4-1)C^3 
	\nonumber \\ 
	&- 34(y_4-2)C^2 + 28(y_4-3)C - 7(y_4-4) \big] \log(1-2C) \Big\}^{-1}. \, 
	\label{eq:k4}
	\end{align}
	%
	Similarly, the gravitomagnetic Love numbers $j_2$, $j_3$ and $j_4$ are given as \cite{Chamel_2021}: 
	\begin{align}
	j_2 =& \, \frac{24}{5} C^5 \big[ 2(\ym_2-2)C - \ym_2 + 3 \big] \times \Big\{ 2C \big[ 2(\ym_2+1)C^3 + 2\ym_2 C^2 + 3(\ym_2-1)C - 3(\ym_2-3) \big] 
	\nonumber \\
	&+ 3 \big[ 2(\ym_2-2)C - \ym_2 + 3 \big] \log(1-2C) \Big\}^{-1} \, , 
	\label{eq:j2} 
	\end{align}
	\vspace{-12pt}
	\begin{align}
	&j_3 = \, \frac{64}{21} C^7 \big[ 8(\ym_3-2)C^2 - 10(\ym_3-3)C + 3(\ym_3-4) \big] 
	\nonumber \\ 
	&\times \Big\{ 2C \big[ 4(\ym_3+1)C^4 + 10\ym_3C^3 + 30(\ym_3-1)C^2 - 15(7\ym_3-18)C + 45(\ym_3-4) \big] 
	\nonumber \\ 
	&+ 15 \big[ 8(\ym_3-2)C^2 - 10(\ym_3-3)C + 3(\ym_3-4) \big] \log(1-2C) \Big\}^{-1} \, , 
	\label{eq:j3} 
	\end{align}
	\vspace{-12pt}
	\begin{align}
	&j_4 = \, \frac{80}{147} C^9 \big[ 40(\ym_4-2)C^3 - 90(\ym_4-3)C^2 + 63(\ym_4-4)C - 14(\ym_4-5) \big]
	\nonumber \\ &\times \Big\{ 2C \big[ 4(\ym_4+1)C^5 + 18\ym_4C^4 + 90(\ym_4-1)C^3 - 5(137\ym_4-334)C^2 
	\nonumber \\ 
	&+ 105(7\ym_4-26)C - 210(\ym_4-5) \big] + 15 \big[ 40(\ym_4-2)C^3 - 90(\ym_4-3)C^2
	\nonumber \\ &+ 63(\ym_4-4)C - 14(\ym_4-5) \big] \log(1-2C) \Big\}^{-1}\, . 
	\label{eq:j4}
	\end{align}
\textls[-33]{	The dimensionless tidal deformability $\Lambda_l$ and $\Sigma_l$ are calculated using \mbox{Equations (\ref{eq:Lambda_l}) and (\ref{eq:Sigma_l})}} for different values of $l$. The Love numbers play a major role in the inspiral phases of BNS merger. The dominant contribution comes mainly from the $l=2$ mode as compared with higher orders polarities, which will be discussed in Subsection \ref{rd:el}. 
	\subsection{Shape/Surficial Love Numbers}
	The surficial Love numbers $h_l$ are associated with the surface deformation of a body. Damour and Nagar gave a relativistic theory of $h_l$ of a body in the influence of tidal forces~\cite{Damour_2009}. It is further extended in Ref. \cite{Landry_2014} to understand it in terms of a deformed curvature of the surface of the body. For a perfect fluid, the surficial Love numbers are related with $k_l$ as $h_l=1+2k_l$ in Newtonian case \cite{Damour_2009, Landry_2014}. We calculate the $h_l$ in the limit of a perfect fluid using the relation given in \cite{Damour_2009, Landry_2014, KumarTide_2017}:
	\begin{equation} 
	h_\ell = \Gamma_1 + 2 \Gamma_2\, k_{\ell}, 
	\label{h_vs_k1} 
	\end{equation} 
	where $\Gamma_1$ and $\Gamma_2$ are defined as 
	\begin{align} 
	\Gamma_1 &= \frac{\ell+1}{\ell-1} (1-M/R) F(-\ell,-\ell;-2\ell;2M/R) 
	- \frac{2}{\ell-1} F(-\ell,-\ell-1;-2\ell;2M/R), \\ 
	\Gamma_2 &= \frac{\ell}{\ell+2} (1-M/R) F(\ell+1,\ell+1;2\ell+2;2M/R) 
	+ \frac{2}{\ell+2} F(\ell+1,\ell;2\ell+2;2M/R), 
	\end{align} 
	The $F(a,b;c;z)$ is the hypergeometric function and 
	$\Gamma_1$ and $\Gamma_2$ can be approximated as follow for $C<<1$ \cite{Landry_2014}:
	\begin{align} 
	\Gamma_1 &= 1 - (\ell+1) (M/R) 
	+ \frac{ \ell(\ell+1)(\ell^2-2\ell+2) }{ (\ell-1)(2\ell-1) } (M/R)^2 
	+ \cdots, \\ 
	\Gamma_2 &= 1 + \ell (M/R) 
	+ \frac{ \ell(\ell+1)(\ell^2+4\ell+5) }{ (\ell+2)(2\ell+3) } (M/R)^2 
	+ \cdots. 
	\end{align}  
	When compactness goes to zero, the surficial Love number attains the Newtonian limit $h_l=1+2k_l$. In the case of a non-rotating black hole, the surficial Love numbers also play a crucial role, as discussed in Refs. \cite{DamourBH_2009, Damour_2009, Landry_2014}. The surficial Love number is independent of the star's internal structure, and this is a function of the mass and radius, i.e., the compactness of the body. 
	\subsection{Tidal Effects on the Inspiral Phase of BNS}
	\label{form:tidal_BNS}
	The GWs emit in the inspiral, merger and post-merger phases of the BNS system. The~BNS in the inspiral phase tidally interacts with each other, and this has a major significance on the waveforms of the GWs. In our earlier \cite{DasMNRAS_2021} study, the DM had substantial effects on the inspiral properties of the BNS. Therefore, in this study, we explore the higher order tidal properties of both the NS with the addition of DM inside it. The~effects of DM on the inspiral properties depend on the factors (i) the type of DM, (ii) the percentage of DM present in the star and (iii) the lifetime of the BNS evolution. In this study, we calculate the effects of tidal corrections to GWs waveforms for DM admixed BNS.
	To calculate the effects of DM on the inspiral/emitted properties of GWs, we use the PN formalism \cite{Blanchet_2006, Boyle_2007, Blanchet_2008, Baiooti_2011, Hotokezaka_2013, Hotokezaka_2016}. The detailed methods can be found in our previous study \cite{DasMNRAS_2021}. We solve the energy balance equation with energy and luminosity up to 3  and 3.5 PN order, respectively. From the energy balance equation, the PN parameter ($x$) is evaluated, which is again used for the calculations of amplitude, phase, etc. as  in Ref. \cite{DasMNRAS_2021}. The~gravitoelectric Love numbers $k_l$ lead to a correction of order ($2l+1$)PN to the phase of the gravitational-wave signal \cite{Yagi_2014, Chamel_2021}:
	\begin{align}
	\Psi_{\ell} &= - \sum_{i=1}^{2} \Biggl[\frac{5}{16}\frac{(2\ell-1)!!(4\ell+3)(\ell+1)}{(4\ell-3)(2\ell-3)} \Lambda_{\ell,i} X_i^{2\ell-1} x^{2\ell-3/2}+ \frac{9}{16}\delta_{\ell 2} \Lambda_{2,i} \frac{X_i^4}{\eta} x^{5/2} \Biggr] 
	\nonumber \\
	&+ \mathcal{O}(x^{2\ell-1/2})\, ,
	\label{eq:Psil}
	\end{align}
	with $X_i=M_i/M$ and $\eta=M_1M_2/M^2$. The $\delta_{\ell\ell'}$ is the Kronecker delta and $i=1,2$ is used to distinguish the two stars of the binary system.
	\section{Results and Discussions}
	\label{R&D}
	In this section, we present our calculated results for the EOS, mass and radius of the NS, various Love numbers and deformabilities in Figures \ref{fig:EOS_mr}--\ref{fig:Sigma}. The tidal deformability is expressed in terms of the gravitoelectric and gravitomagnetic Love numbers, and the surface deformation of the star is described by the surficial Love numbers. The detailed discussions are outlined in the subsequent subsections.
	\subsection{Equation of State and Mass-Radius Relations}
\begin{figure}[H]
\includegraphics[width=0.8\textwidth]{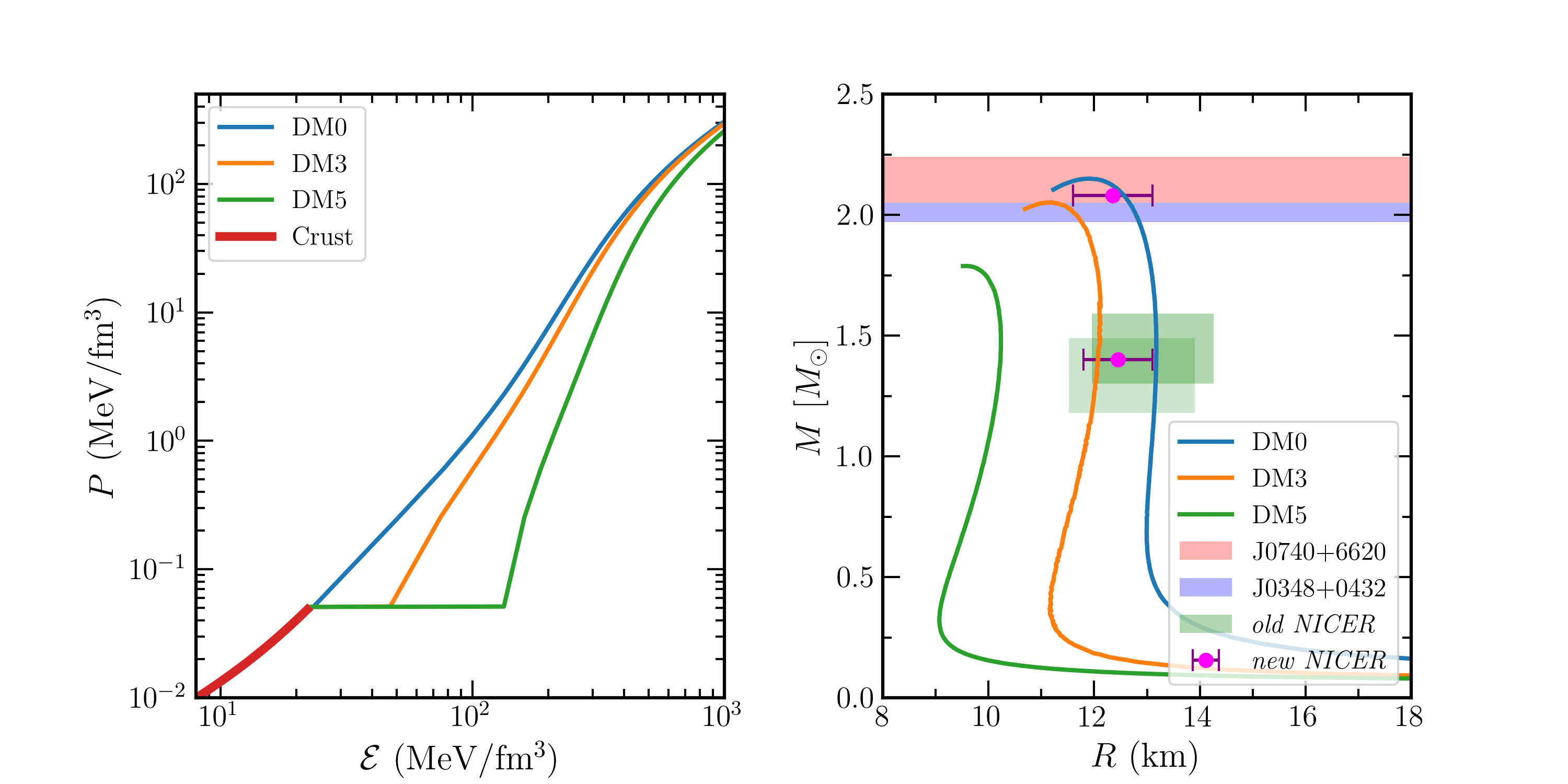}
		\caption{{ (\textbf{Left}):} EOSs for unified IOPB-I (IOPB-I-U) with different DM percentages. The red line represents the IOPB-I crust calculated in Ref. \cite{Parmar_2021}. { (\textbf{Right}): } $M-R$ profiles for DM admixed NS. Different massive pulsars constraints, such as MSP J0740+6620 by Cromartie et al. \cite{Cromartie_2020} and PSR J0348+0432 by Antoniadis et al. \cite{Antoniadis_2013} are overlaid with different colour bars. Both old NICER \cite{Miller_2019, Riley_2019} constraints for canonical stars and new NICER limits for both 1.40 and 2.08 $M_\odot$ are also depicted \cite{Miller_2021}.}
		\label{fig:EOS_mr}
\end{figure}

	The IOPB-I unified EOS (IOPB-I-U), represented as DM0, and two DM admixed EOSs, such as DM3 and DM5 (DM0, DM3 and DM5 represent the DM Fermi momenta 0.00, 0.03 and 0.05 GeV, respectively.), are shown on the left side of Figure \ref{fig:EOS_mr}. At the lower density region, all three EOSs coincide with each other in to single line represents the crust part as shown in the figure. This means the crust EOS is the same for all three cases.  With~the addition of DM, the EOSs become softer, which is clearly seen in the figure. The softening of the EOS depends on the amount of DM contained inside the NS. Therefore, EOS with 0.05 GeV DM momentum is the softest as compared to the other two cases, DM3 and DM0.
	
On the right side of Figure \ref{fig:EOS_mr}, we depict the mass $M$ and radius $R$ of the DM admixed NS with the IOPB-I-U equation of state. The IOPB-I-U predicts a mass of $M=2.149 \ M_\odot$ and radius $R=11.748$ km. Both the mass and radius are  well within the known observational constraints, as shown in the figure. With the addition of DM, the $M$ and $R$ of the NS decreases as the presence of DM softens the EOSs. {We noticed that there is a certain range for DM momentum produces mass--radius values compatible with the latest massive pulsar and NICER limit. Hence, from the observational data, such as advance LIGO and NICER, one can fix the percentage of DM inside the NS. Theoretically, by generating many such EOSs and using Bayesian analysis, one could also fix the DM percentage inside the NS.}
	\subsection{Electric Love Number and Tidal Deformability}
	\label{rd:el}
    \begin{figure}
		\includegraphics[width=0.8\textwidth]{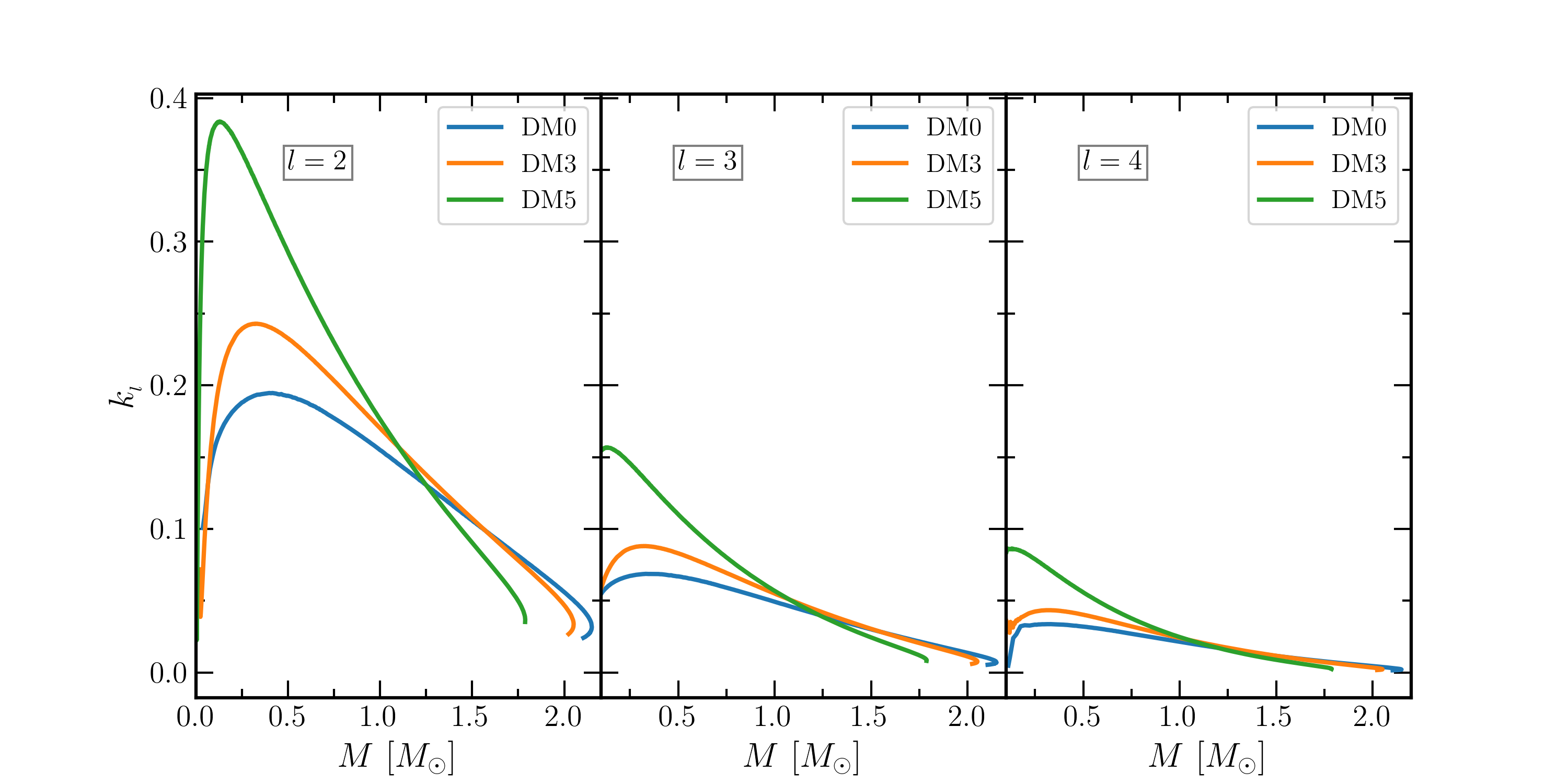}
		\caption{Gravitoelectric tidal Love numbers for IOPB-I-U EOS with DM Fermi momenta 0.00, 0.03 and 0.05 GeV with $l=2, 3, {\rm and } \  4 $.}
		\label{fig:kl}
	\end{figure}
\begin{figure}
		\includegraphics[width=0.8\textwidth]{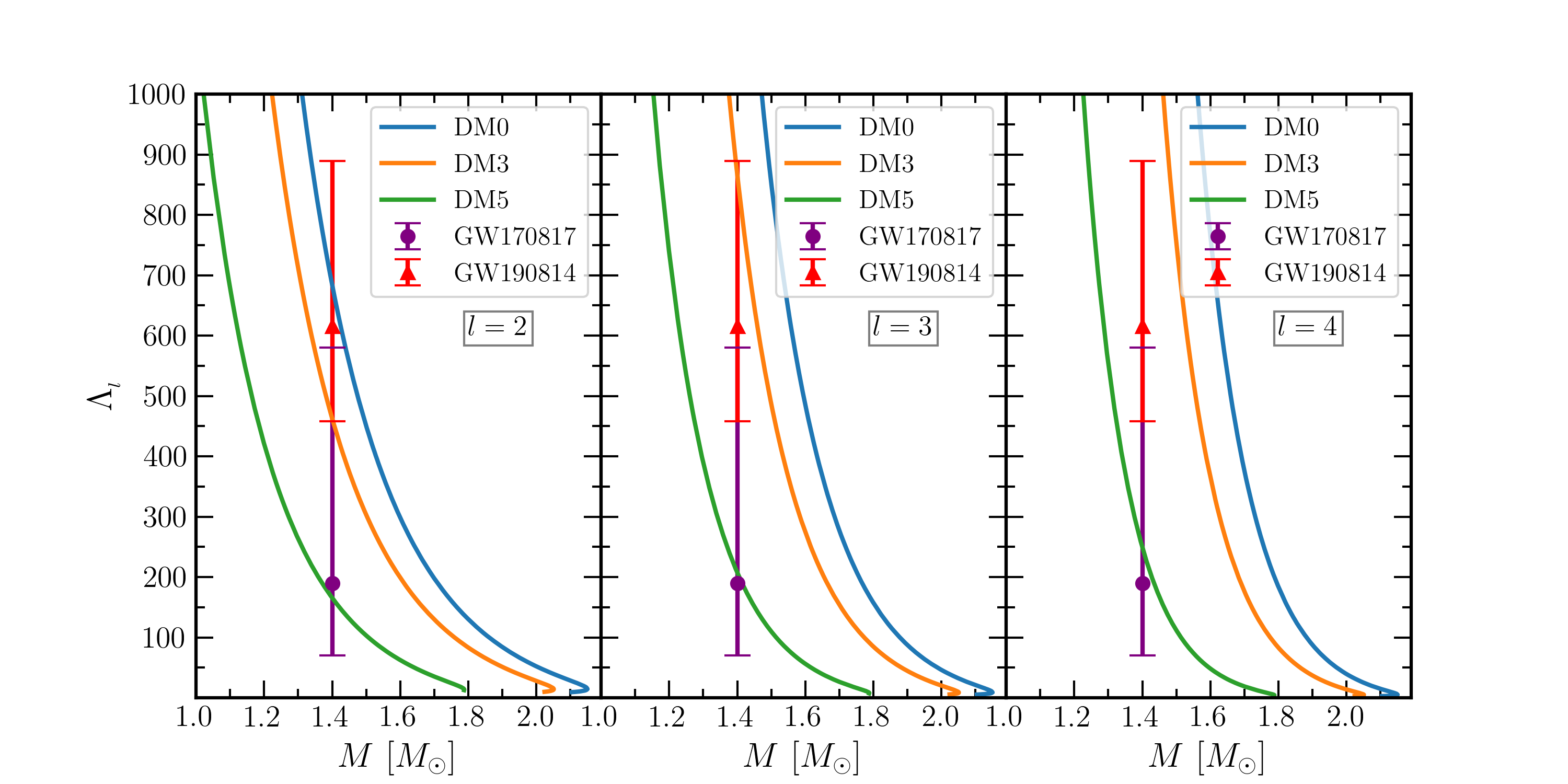}
		\caption{Same as Figure \ref{fig:kl} but for dimensionless electric tidal deformability. The purple colour error bar represents the constraint on $\Lambda_{1.4}$ given by LIGO/Virgo \cite{Abbott_2017, Abbott_2018} from the BNS merger event GW170817 with, $\Lambda_{1.4}=190_{-70}^{+390}$. The red colour error bar represents the $\Lambda_{1.4}$ constraints from the GW190814 event under the assumption of NSBH scenario with, $\Lambda_{1.4}= 616_{-158}^{+273}$ \cite{RAbbott_2020}.}
		\label{fig:Lambda}
	\end{figure}

	The gravitoelectric Love numbers with different values of $l$ are calculated using Equations (\ref{eq:k2})--(\ref{eq:k4}), and the results are shown in Figure \ref{fig:kl}. As  expected, the value of $k_2$ is more pronounced than the $k_3$ and $k_4$. The effects of crust on the $k_l$ are more significant as compared to the EOS without taking the crust into consideration as observed in Ref. \cite{Chamel_2020}. Hence, we take the unified EOSs to calculate the $k_l$. The effects of DM on the gravitoelectric Love numbers are also shown in the figure. The maximum value of $k_l$ increases with DM percentage, and the differences are seen mainly in the lower mass regions, where the crust plays an important role. Hence, it is crucial to choose  suitable crust EOSs for Love numbers and tidal deformability calculations.
	With the increase of $l$, the magnitude of $k_l$ decreases in a more profound manner. However, the impacts of DM on $k_3$ and $k_4$ are non-negligible. It is discussed in Ref. \cite{Chamel_2021} that the symmetry energy has less importance on the crust compared with the stiffness of the EOS. We also found similar results for the DM admixed NS, as the EOSs softens with the addition of DM.
	The dimensionless tidal deformability $\Lambda_l$ were calculated for DM admixed NS within the IOPB-I-U equation of state, and the results are shown in Figure \ref{fig:Lambda}. All the three EOSs, such as DM0, DM3 and DM5, well satisfy the $\Lambda_{1.4}$ constraints for $l=2$. On the other hand, we found a shifting on the $\Lambda_l$ with the increased $l$ values, which do not satisfy the GW170817 limit except for higher DM momenta.
	
	The secondary component of the GW190814 event is very much in debate, ``{\it whether it is a supermassive NS or the lightest black hole?}". Several attempts have been taken to explore this object \cite{Huang_2020, DasPRD_2021}. Some of the suggestions are as follows: (i) It is a heavy NS with deconﬁned QCD core \cite{Tan_2020}, (ii) a super-fast pulsar \cite{Zhang_2020}, (iii) a binary black-hole merger \cite{Fattoyev_2020, DasBig_2021} and (iv) the DM admixed NS with EOS is sufficiently stiff \cite{DasPRD_2021} etc. If this event is the merger of the NS and black hole (NSBH), then the tidal deformability of the NS for the canonical star is $\Lambda_{1.4}= 616_{-158}^{+273}$. 
	Hence, one can impose a constraint on the NS mass, radius and tidal deformability from both GW170817 and GW190814 data. {We noticed that, using IOPB-I EOS,  DM0 and DM3 satisfy the values of $\Lambda_l$ given by GW190814 for the lowest multipole moment. For higher values of $l$, the values of $\Lambda_l$ are shifted and do not satisfy the GW190814 limit.}
	\subsection{Magnetic Love Number and Tidal Deformability}
\begin{figure}
		\includegraphics[width=0.8\textwidth]{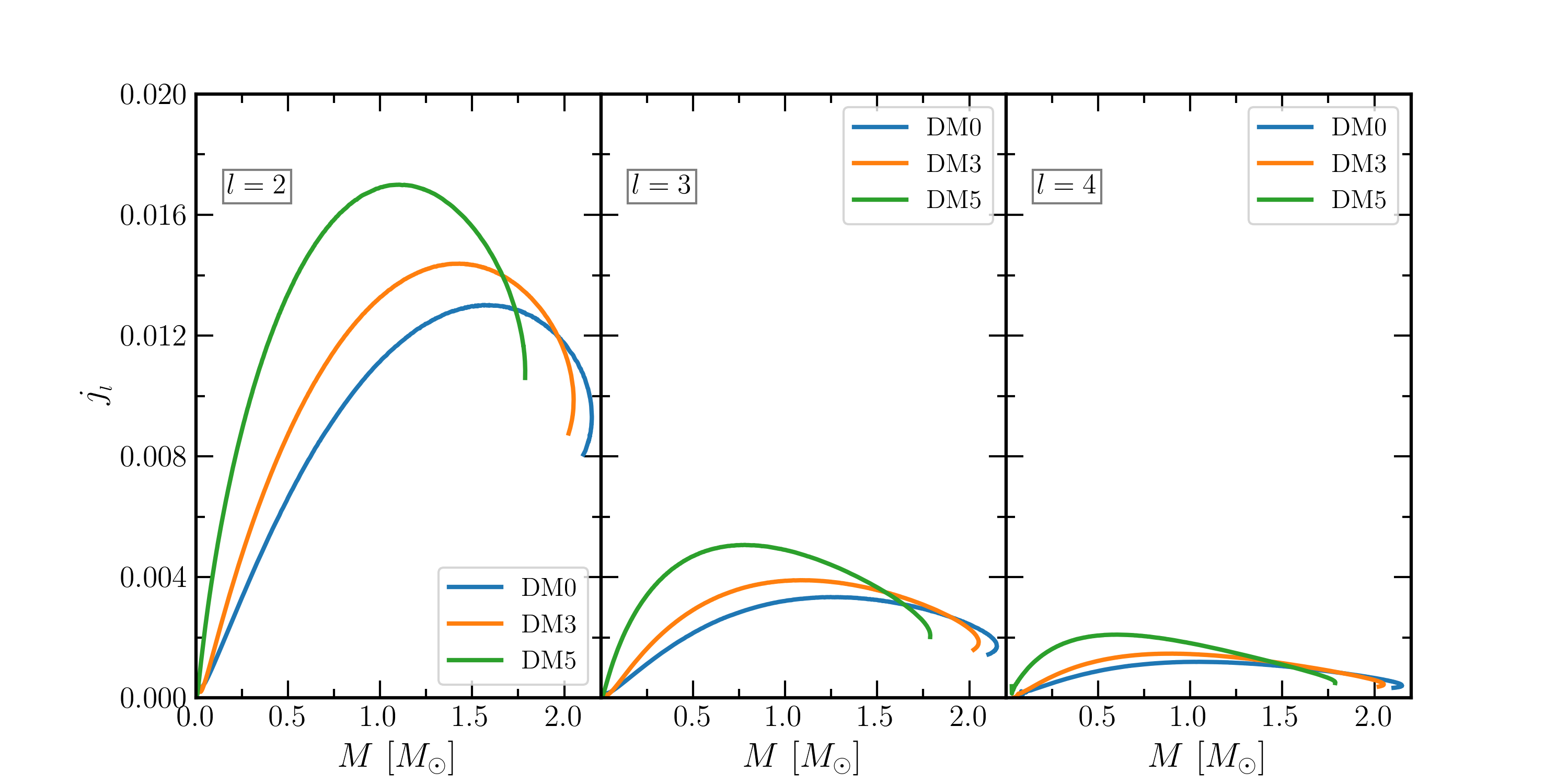}
		\caption{Magnetic tidal Love numbers for IOPB-I EOS with DM Fermi momenta 0.00, 0.03 and 0.05 GeV with three different $l$ values for static fluid.}
		\label{fig:jl_static}
	\end{figure}

\begin{figure}
		\includegraphics[width=0.8\textwidth]{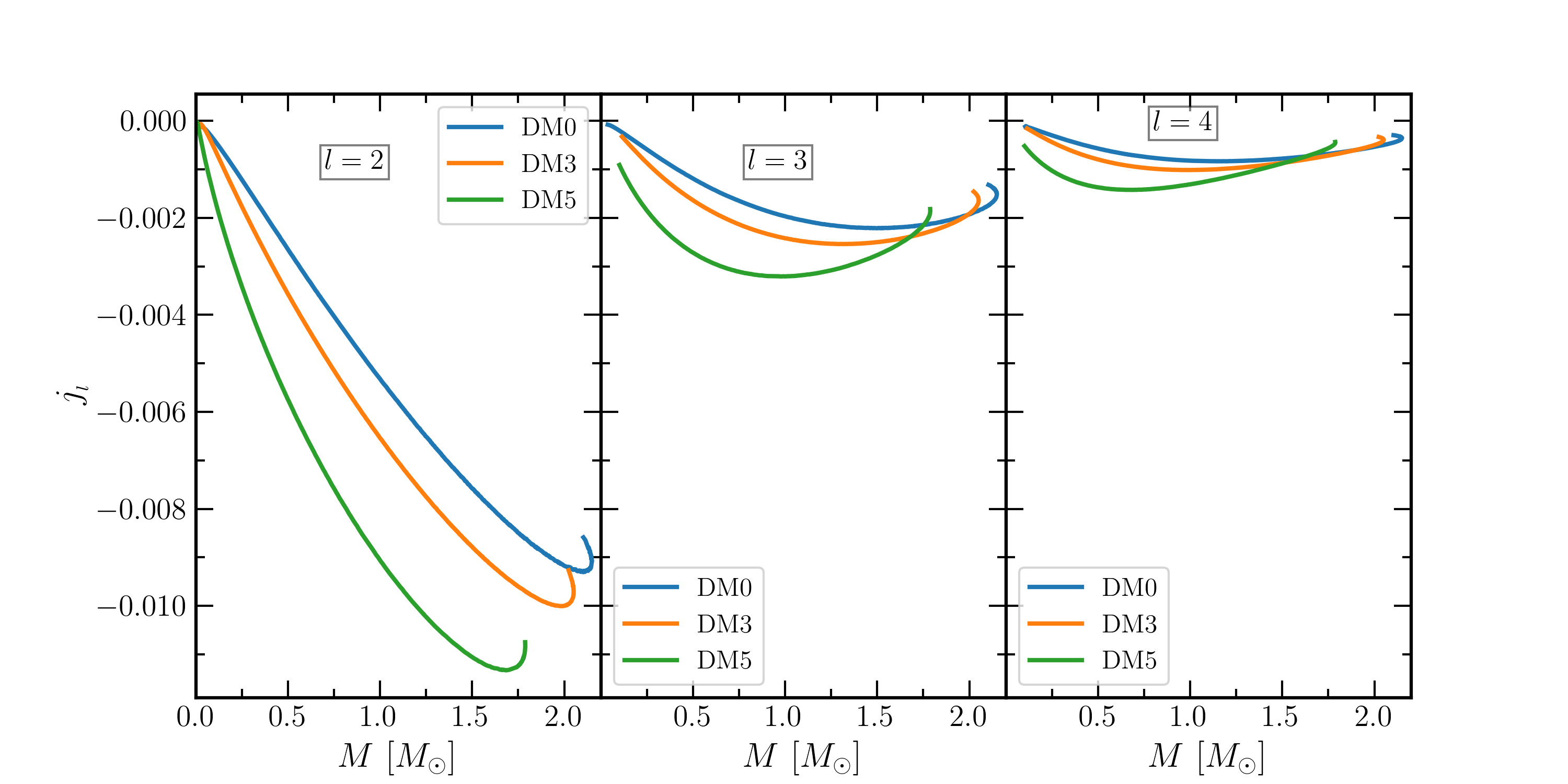}
		\caption{Same as Figure \ref{fig:jl_static} but for irrotational fluid.}
		\label{fig:jl_irrot}
	\end{figure}

\begin{figure}
		\includegraphics[width=0.8\textwidth]{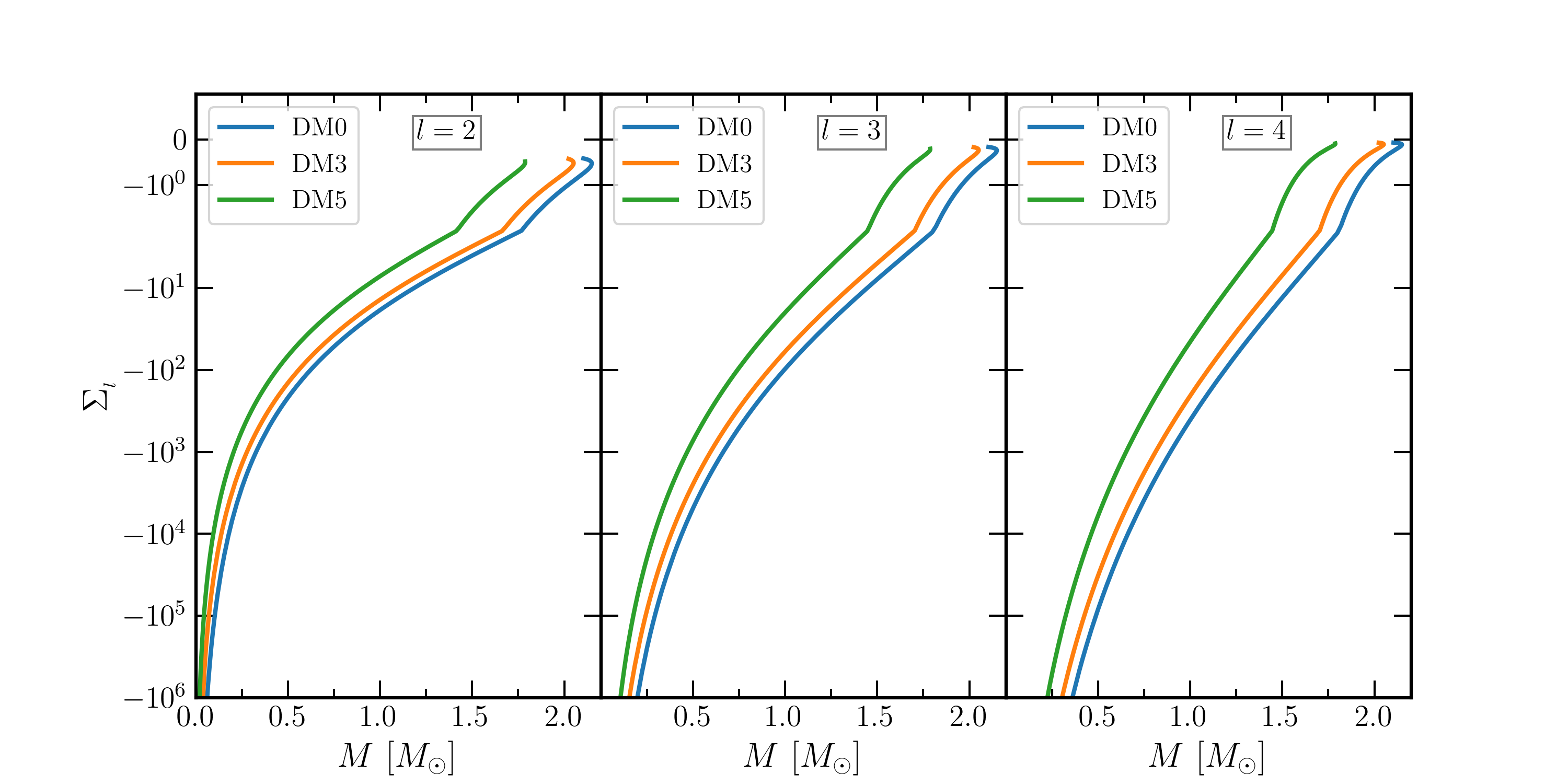}
		\caption{Same as Figure \ref{fig:jl_irrot} but for dimensionless magnetic tidal deformability.}
		\label{fig:Sigma}
	\end{figure}

	We calculate the gravitomagnetic Love numbers and tidal deformabilities for \mbox{$l=2, 3, 4$} using Equations (\ref{eq:Sigma_l}) and (\ref{eq:j2})--(\ref{eq:j4}), and the results are depicted in Figures \ref{fig:jl_static} and  \ref{fig:jl_irrot}. The~magnitude of $j_2$ is more negative as compared to both $j_3$ and $j_4$, i.e.,  with the addition of DM in the EOSs, the $j_l$ values correspond to DM3 and DM5 decrease as compared to DM0 case. Hence, effects of DM on the $j_l$ are significant. This also depends on the DM contained inside the NS. The magnitudes of the gravitomagnetic tidal Love numbers for both static and irrotational fluids are almost comparable; however,  their values are in opposite signs (as clearly seen from figures). These tidal Love numbers also impinge less effect on the waveform of GWs, which might be detected in the upcoming modern detectors. 
	The gravitomagnetic tidal deformability is shown in Figure \ref{fig:Sigma} for DM admixed NS. The values of $\Sigma_l$ increase with DM momenta, contrary to the $\Lambda_l$. However, this effect is  less significant on the waveforms of the emitted GWs since it depends on the value of $j_l$ as compared to $\Lambda_l$. There is also no significant changes with different values of $l$ with the DM percentage. We also observed that the changes are seen mainly in the core part of the NS, as the crust has less contribution to $j_l$ and $\Sigma_l$. Hence, the crust part does not affect both the gravitomagnetic Love numbers and tidal deformabilities.
	\subsection{Surficial Love Numbers}
\begin{figure}
		\includegraphics[width=0.8\textwidth]{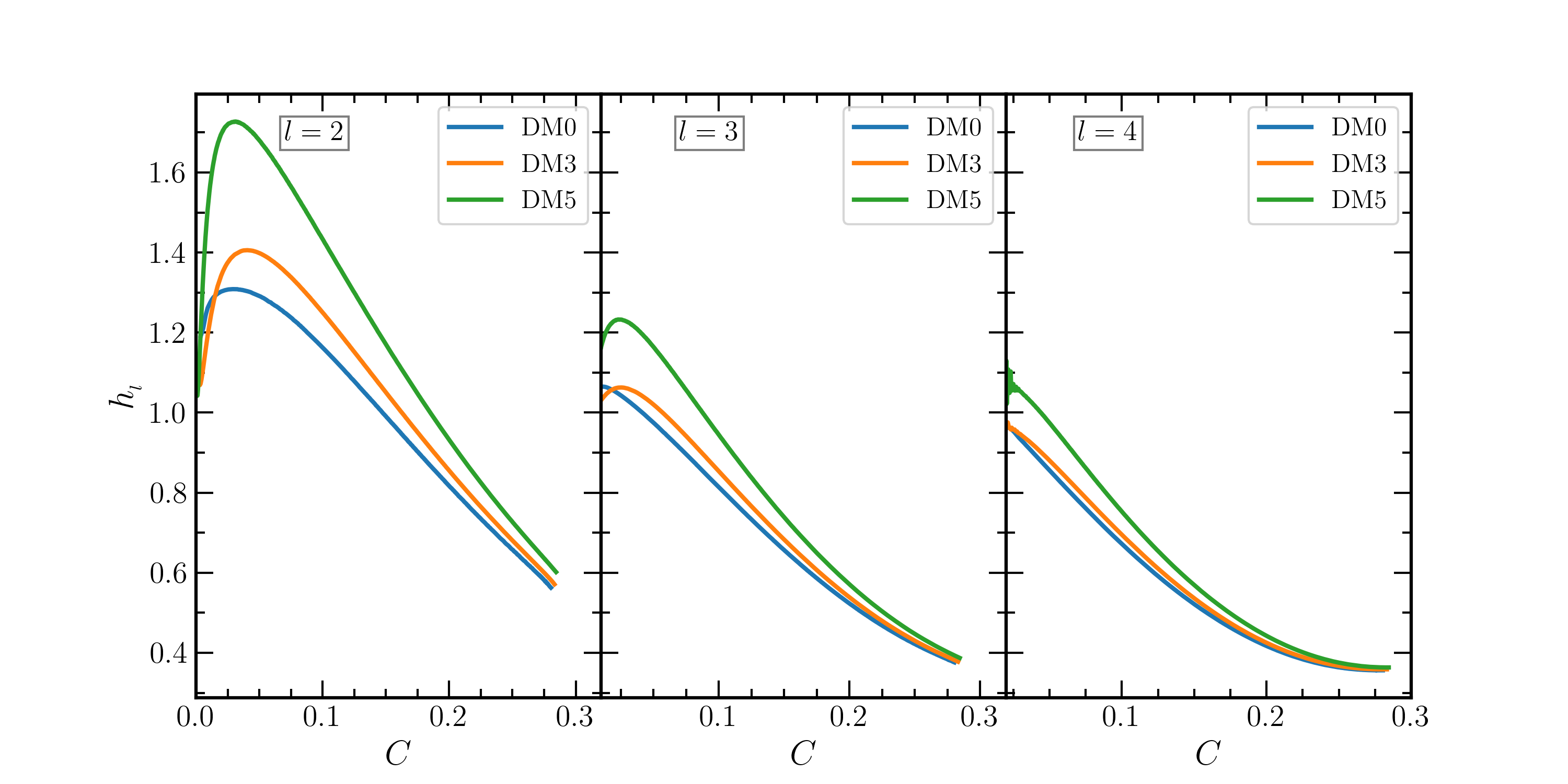}
		\caption{Shape/surficial Love numbers for IOPB-I EOS with DM Fermi momenta 0.00, 0.03 and 0.05 GeV with three different $l$.}
		\label{fig:hl}
	\end{figure}

	The surficial Love numbers describe the surface deformation of a body in the presence of the tidal field. These are expressed in terms of the compactness of the NS in this calculations. The calculated surficial Love numbers $h_l$ with DM admixed NS are shown in Figure \ref{fig:hl} for different multipole moments $l$. The pattern of the variation of $h_l$ is almost similar to the $k_l$, but the magnitudes $h_l$ is  high compared to $k_l$. When the compactness $C$$\rightarrow$$0$, all the values of $h_l$, i.e., $h_2$, $h_3$ and $h_4$  approach  $\sim{1}$, irrespective of the moments (see Figure \ref{fig:hl}). 
%
	
	The value of $h_l$ increases with the percentage of DM contained in the NS. Thus, the surface of the NS is more distorted for the DM5 case than the other two DM3 and DM0 conditions. A comparative analysis among the multipole moments tells that the surface deformation is mainly due to the quadrupole deformation of the star. However, the contributions of $h_l$ for $l=3$ and 4 are non-negligible. In all the cases of $l$, the maximum distortion of the star takes place at around $C\sim{0.04}$. 
	\subsection{Tidal Effects on the GW Signal}
\begin{figure}
		\includegraphics[width=0.8\textwidth]{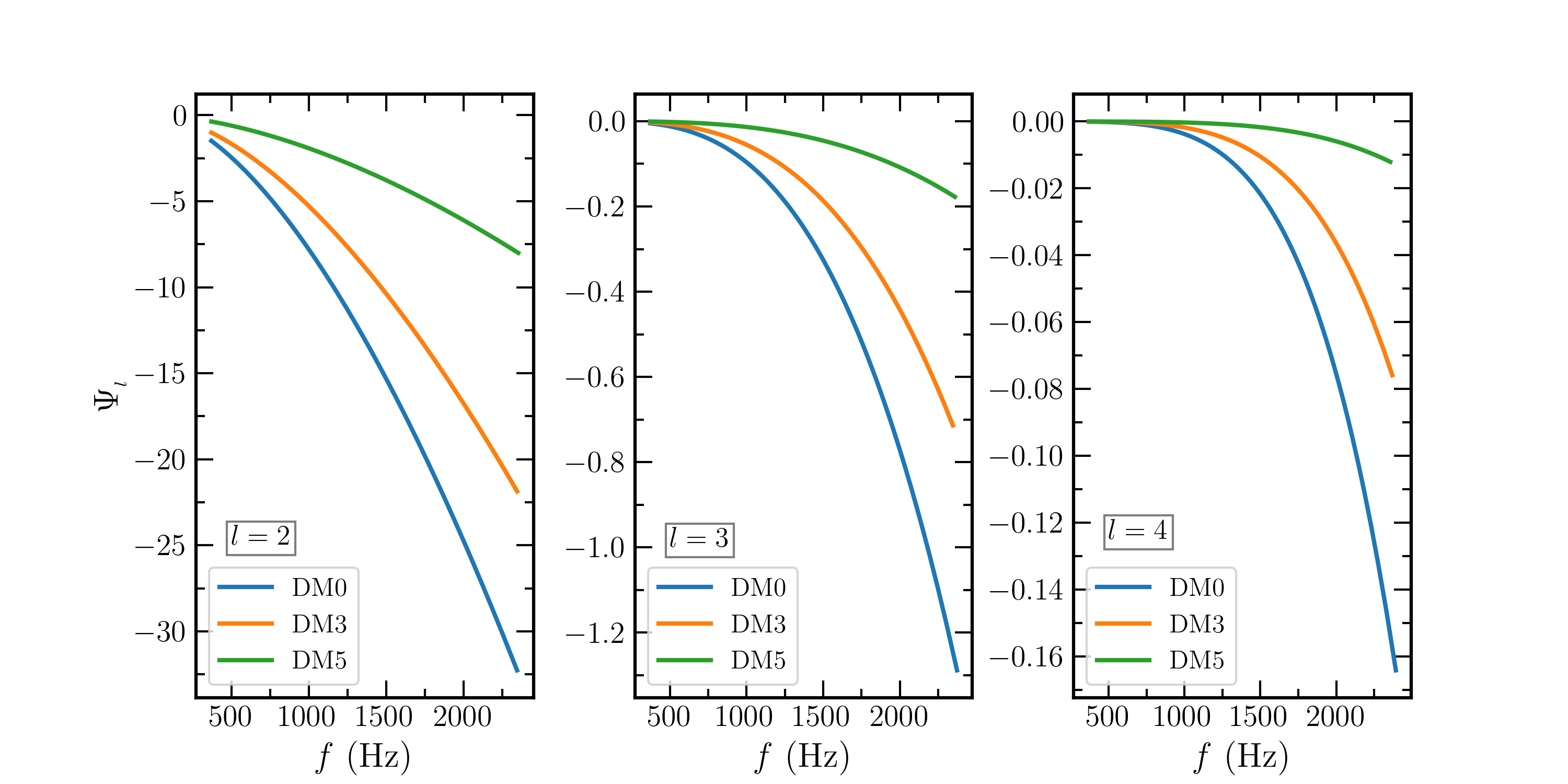}
		\caption{Tidal corrections as a function of gravitational frequency for IOPB-I EOS with DM Fermi momenta 0.00, 0.03 and 0.05 GeV with three different $l$.}
		\label{fig:psil}
	\end{figure}

	The tidal deformations have significant effects on the gravitational wave in the later stage of the inspiral phase. In this calculation, we  study the effects of DM on tidal corrections on the gravitational wave. The tidal corrections are calculated using a PN approximation. The correction is only related with NS observable through the tidal deformability parameter, which is model dependent. In our previous study \cite{DasMNRAS_2021}, we  calculated the in-spiral properties with a DM equation of state, and we observed that the tidal deformability of the star decreased with increasing DM percentage, which slows down the merger dynamics. The effects of DM on the GW amplitude (such as $h_+$, $h_\times$ and $h_{22}$), inspiral phase ($\Phi$) etc. are calculated. Here, we calculate the tidal corrections ($\Psi_l$) and depict these in Figure~\ref{fig:psil} for different values of $l$ as a function of frequency ($f$). Similar to all previous cases, we observe that the leading order contribution comes from the $l=2$ case. The magnitude of the tidal contribution was found to be greater in the higher frequency region. Hence, the~corrections also accelerate the merger of the BNS for a higher massive star than the less massive one.
The mass of the NS decreases with the addition of DM to the BNS. From our calculated results (see Figure \ref{fig:psil}), it is clear that the magnitude of $\Psi_l$ decreases with the increasing percentage of DM. We can connect this behaviour with the decreasing nature of $\Lambda_l$, as $\Psi_l$ is related with $\Lambda_l$ (see Equation (\ref{eq:Psil})). Thus, the merging time of the binary system increases with the addition of DM in the NSs. Similar to other quantities, the contribution to waveform $\Psi_l$ has less significance from the higher-order deformation of the binary system. Thus, most of the GWs signal comes from the quadrupole deformations.
	\section{Conclusions}
	\label{conclusion}
	In summary, we calculated the macroscopic properties of NSs, such as the mass, radius and compactness, by solving the TOV equations as an input of the effective field-theory-motivated relativistic mean-field equation of states. The recently developed IOPB-I-U unified EOS was used in the calculations. The obtained results are well within the recent pulsars and NICER observational data. 
	Then, the gravitoelectric and gravitomagnetic tidal Love numbers along with the dimensionless deformability were calculated for higher multipole moments with DM admixed NS. We found significant effects of the DM on both the electric and magnetic deformation properties of the BNS inspiralling system. We also estimated the surficial deformability for various multipole moments, and similar to the tidal deformability, the contribution of the $h_l$ was the maximum for $l=2$ followed by higher moments. These $h_l$ contributions mainly depend on the compactness of the star. 
    The effects of DM on the properties of the NS were investigated by varying the DM percentage/Fermi momenta. The values of gravitoelectric and gravitomagnetic Love numbers increased with increasing DM Fermi momenta for all values of $l$ contrary to tidal deformability. The gravitoelectric tidal deformability was well satisfied with the observational limits imposed by the GW170817 event (in the quadrupole case). 
    
    With the increasing values of multipole moments---i.e., for higher $l$---the dimensionless deformability deviated from the GW170817 observational data for DM0 and DM3; however,  the calculated deformability with DM5 still satisfied the data for $l=3, \ {\rm and} \ 4$. Similar to the GW170817 limit, the secondary component of the GW190814 event also put a limit on the dimensionless tidal deformability for the NS-BH merger. 
    NS with higher DM percentages, such as DM3 and DM5, satisfy the GW190814 constraints for higher multipole-moments. Up to a certain DM density, the~predicted NS properties satisfy the observational data. From the observational data, one can also fix the DM percentage by generating many such EOSs and using Bayesian analysis.
    We also calculated the tidal corrections on the emitted GWs in the inspiral phases of the binary system. Using the PN formalism, the tidal corrections for different values of $l$ were calculated for DM-admixed BNS. The magnitude of the tidal correction decreased with the addition of DM in the NS equation of state. For higher multipole moments (larger $l$), the corrections were almost negligible as compared to $l=2$. In our previous study \cite{DasMNRAS_2021}, we observed that the DM slowed down the merger of the BNS. This is because the values of the macroscopic properties are reduced with increasing DM percentages resulting in smaller tidal corrections compared with the NS without  DM components.  
	
    In the present calculations, we focused on studying the BNS in the inspiral phase and various deformabilities. The merger and post-merger dynamics of the DM admixed BNS system are beyond the scope of this calculation. It is still in debate ``{\it whether post-merger remnant is  a hyper-massive NS or a black hole?''} \cite{Sarin_2021}. The post-merger remnant of the DM-admixed BNS merger depends on the amount and type of DM present inside merger systems. In~\mbox{Refs. \cite{Pollack_2015, Ellis_2018}}, it was reported that, if the DM is a self-interacting type, an extra peak should be seen in the post-merger frequency template/power spectral density. Therefore, we suggest  including the DM inside the NS while modelling the inspiral-merger-ringdown templates for the GWs detection.
	\vspace{6pt} 
	
	\authorcontributions{Conceptualization, H. C. Das, Bharat Kumar and S. K. Patra; data curation, H. C. Das and Ankit Kumar; formal analysis, H C Das and Ankit Kumar; investigation, H. C. Das and Bharat Kumar; writing-original draft, H. C. Das, Bharat Kumar and S. K. Patra; writing-review and editing, Bharat Kumar and S. K. Patra. All authors have read and agreed to the published version of the manuscript}

\funding{Not applicable.} 

\institutionalreview{Not applicable.} 

\informedconsent{Not applicable} 

\dataavailability{The data presented in this study are available on request from the corresponding author.} 

\acknowledgments{H.C.D. would like to thank P. Landry and N. Chamel for the discussions on the tidal Love numbers.}

\conflictsofinterest{The authors declare no conflict of interest.} 
\end{paracol}
\bibliography{dm.bib}
\bibliographystyle{mdpi}
 	\end{document}